\documentstyle[12pt]{article}
\newcommand{\bibi}{\bibitem}

\newcommand{\un}{1\!\!1}

\newcommand{\al}{\alpha}

\newcommand{\lag}{\langle}
\newcommand{\rag}{\rangle}
\newcommand{\gm}{\gamma}

\newcommand{\Gm}{\Gamma}
\newcommand{\dl}{\delta}
\newcommand{\ep}{\varepsilon}

\newcommand{\vep}{\varepsilon}
\newcommand{\vst}{v^{st}}
\newcommand{\fst}{f^{st}}

\newcommand{\zt}{\zeta}
\newcommand{\et}{\eta}

\newcommand{\kp}{\kappa}
\newcommand{\lm}{\lambda}

\newcommand{\vphibar}{\overline{\varphi}}

\newcommand{\rh}{\rho}
\newcommand{\sg}{\sigma}

\newcommand{\ph}{\phi}
\newcommand{\vr}{\varphi}

\newcommand{\ps}{\psi}

\newcommand{\Ps}{\Psi}

\newcommand{\Psb}{\overline{\Ps}}
\newcommand{\psb}{\overline{\ps}}

\newcommand{\mh}{\hat{\mu}}

\newcommand{\chb}{\overline{\chi}}
\newcommand{\hmu}{\hat{\mu}}
\newcommand{\phat}{\hat{p}^2}
\newcommand{\rhh}{\hat{\rho}}

\newcommand{\half}{\mbox{{\small $\frac{1}{2}$}} }
\newcommand{\third}{\mbox{{\small $\frac{1}{3}$}} }

\newcommand{\eighth}{\mbox{{\small $\frac{1}{8}$}} }

\newcommand{\Sm}{\sum_{\mu}}
\newcommand{\Tr}{\mbox{Tr}}
\newcommand{\Det}{\mbox{Det}}
\newcommand{\del}{\partial}
\newcommand{\dm}{\del_{\mu}}
\newcommand{\dg}{\dagger}
\newcommand{\ra}{\rightarrow}

\newcommand{\be}{\begin{equation}}
\newcommand{\ee}{\end{equation}}
\newcommand{\bea}{\begin{eqnarray}}
\newcommand{\eea}{\end{eqnarray}}
\newcommand{\eq}{\ref}
\newcommand{\beq}{\begin{equation}}
\newcommand{\eeq}{\end{equation}}
\newcommand{\cc}{\cite}
\newcommand{\lb}{\label}
\newcommand{\gsim}{\stackrel{>}{\sim}}  

\def \3{\ss}

\def\dateandnumber(#1)#2#3#4{
\vbox to 18mm{%
     \hbox to \textwidth{ \hspace*{14mm} \hsize=40mm%
            \vbox{%
                 \hbox to 40mm{\large #1 \hss}%
                 \hbox to 40mm{    \hss}%
                 \hbox to 40mm{    \hss}%
                 }%
                 \hss \hsize=80mm%
            \vbox{%
                 \hbox to 80mm{\hss \large #2}
                 \hbox to 80mm{\hss \large #3}
                 \hbox to 80mm{\hss \large #4}
                 }%
            \hspace*{14mm} }%
      \vss
    }
}
\def\titleofpreprint#1#2#3#4{{\LARGE \bf
\vbox to 43mm{%
     \vss
     \hbox to \textwidth{ \hspace*{14mm} \hsize=130mm%
            \hss \vbox{
                      \hbox to 130mm{\hss \LARGE \bf #1\hss}%
                      \hbox to 130mm{\hss \LARGE \bf #2\hss}%
                      \hbox to 130mm{\hss \LARGE \bf #3\hss}%
                      \hbox to 130mm{\hss \LARGE \bf #4\hss}%
                 }%
            \hss \hspace*{14mm} }%
      \vss
    }}
}
\def\listofauthors#1#2#3{{\large
\vbox to 22mm{%
     \vss
     \hbox to \textwidth{ \hspace*{14mm} \hsize=130mm%
            \hss \vbox{
                      \hbox to 130mm{\hss \large #1\hss}%
                      \hbox to 130mm{\hss \large #2\hss}%
                      \hbox to 130mm{\hss \large #3\hss}%
                 }%
            \hss \hspace*{14mm} }%
      \vss
    }}
}
\def\listofaddresses#1#2#3{{\small
\vbox to 18mm{%
     \vss
     \hbox to \textwidth{ \hspace*{14mm} \hsize=130mm%
            \hss \vbox{
                      \hbox to 130mm{\hss \small #1\hss}%
                      \hbox to 130mm{\hss \small #2\hss}%
                      \hbox to 130mm{\hss \small #3\hss}%
                 }%
            \hss \hspace*{14mm} }%
      \vss
    }}
}
\def\abstractofpreprint#1{{\normalsize
\vbox to 110mm{%
     \vss
     \hbox to \textwidth{\hss \normalsize \bf Abstract \hss}%
     \vspace*{1cm} \normalsize
     #1
     \vss
    }}
}

\def\footnoteitem(#1)#2{
\begin{list}{#1}{\labelwidth4.0mm \leftmargin7.0mm
\labelsep2.5mm \rightmargin7.0mm \parsep0.5ex plus0.2ex minus0.1ex
\itemsep0ex plus0.2ex }
\item #2
\end{list}
}
\begin{document}
\dateandnumber(April 1992)%
{Amsterdam ITFA 92-13}%
{J\"ulich HLRZ 92-23}%
{                    }%
\titleofpreprint%
{      Fermion-Higgs model with                                }%
{      Reduced Staggered Fermions                              }%
{                                                              }%
{                                                              }%
\listofauthors%
{Wolfgang~Bock$^{1,\$}$, Jan Smit$^{1,*}$ and Jeroen C. Vink$^{2,\#}$}%
{                                                              }%
{                                                              }%
\listofaddresses%
{\em $^1$Institute of Theoretical Physics, University of Amsterdam,}%
{\em \\ Valckenierstraat 65, 1018 XE Amsterdam, The Netherlands}%
{\em $^2$HLRZ c/o KFA J\"ulich,                                 %
     PO Box 1913, D-5170 J\"ulich, Germany                    }%
\abstractofpreprint{
We introduce a lattice fermion-Higgs model with
one component `reduced staggered' fermions. In order to use
the fermion field as efficiently as possible we couple the two
{\em staggered} flavors to the O(4) Higgs field leading to a model with
only one SU(2) doublet in the scaling region. The number of fermions is
doubled in a numerical investigation
of the model with the hybrid Monte Carlo algorithm. We present
results for the phase diagram, particle masses and
renormalized couplings on lattices ranging in size from $6^3 24$ to $16^3 24$.
}
\vspace{1cm}
\begin{flushleft}
$\$$ e-mail: bock@phys.uva.nl   \\
$*$ e-mail: jsmit@phys.uva.nl   \\
$\#$ e-mail: vink@hlrsun.hlrz.kfa-juelich.de
\end{flushleft}
\thispagestyle{empty}
\setcounter{page}{0}
\pagebreak
\subsection*{1. Introduction}
Recent years have witnessed a lot of interest in the non-perturbative
understanding of the symmetry breaking sector of the standard model.
An important question
is whether the scalar field self-coupling and the Yukawa coupling
are trivial as suggested by the signs of the perturbative $\beta$
functions. If the model is influenced by the Gaussian fixed point
for all allowed values of the bare couplings,
one can derive non-perturbative upper bounds
on the renormalized Yukawa coupling $y_R$ and
quartic self-coupling $\lambda_R$
from the constraint that the cut-off must be sufficiently larger than the
masses of the particles. Using the relations
$m_H = \sqrt{2 \lm_R} v_R$ and $m_F = y_R v_R$
this is equivalent to finding upper bounds on the Higgs mass $m_H$
and the heavy fermion mass $m_F$ ($v_R \approx 246$ GeV
is the scalar field expectation value).

It is reasonable to address these questions first
in simplified lattice fermion-Higgs models without gauge fields.
For recent reviews
on such models we refer the reader to ref.~\cc{REV}.

In a model with naive fermions
the number of mass degenerate doublets is as large as 16.
With the usual staggered fermions one can reduce this number of doublets
to four and using the mirror fermion model with Wilson fermions
of ref.~\cc{Mo87} it appears to be
possible to reduce the number of doublets to one, while the
30 doublers acquire masses of order of the cut-off \cc{Mo92}.
For numerical investigations of these models with the hybrid Monte Carlo
algorithm (HMCA) it is necessary to double these numbers of fermion doublets.

In this letter we shall investigate a new proposal \cc{Sm91}
for a fermion-Higgs model which is based on the
reduced staggered fermion formalism and which describes one SU(2) doublet in
the
scaling region (two doublets when investigating the model with the HMCA).
In this paper we shall introduce the model and present
some preliminary results. A detailed account of our investigation
will appear elsewhere \cc{BoFr92a}.
%
%
%
%
\subsection*{2. The model and its symmetries }
The usual euclidean staggered fermions on a four-dimensional hypercubic lattice
describe four flavors in the scaling region.
By using the `reduced' staggered formalism \cc{ShTh81,DoSm83}
this number can be halved to two. If two of these staggered fields are placed
in an SU(2) doublet one would get a model
with two degenerate isospin doublets in the
scaling region. In this letter, however, we follow another strategy. In order
to use the fermion field as efficiently as possible
we want to couple the two {\em staggered}
flavors to the Higgs field, leading to a model with only one doublet
in the scaling region.

To show how the Higgs field can be coupled to the staggered
fermion flavors while preserving as much symmetry as possible,
we first write down the target action that we want to
reproduce with the lattice model,
\be
 S_F   =   -\int d^4x [\psb  \gm^{\mu} \partial_{\mu} \ps
         +  y (\psb_L \ph \ps_R + \psb_R \ph^{\dg} \ps_L )]  \lb{SFC}
\ee
The Dirac doublet $\ps$ interacts with the
scalar field $\ph \in SU(2)$ which contains the O(4) components
$\vr_{\mu}$ of the Higgs field in the familiar way,
$\ph =  i \sum_{j=1}^{3} \varphi_j \tau_j + \varphi_4 \un $
where $\tau_{j}$ are the usual Pauli matrices.
%
%

In order to reproduce this action with staggered fermions, we need to
exhibit their spin-flavor structure and
it is convenient to introduce the $4 \times 4$
matrix fields \cc{Sm88}
\be
\Psi_x=\eighth\sum_{b}  \gm^{x+b}       \chi_{x+b} ,\;\;\;
\Psb_x=\eighth\sum_{b} (\gm^{x+b})^{\dg}\chb_{x+b} \lb{SDT} .
\ee
Here $\chi$ and $\chb$ are the usual (one-component) staggered
fields, the sums in eq.~(\eq{SDT}) are over the 16 corners of the unit
hypercube, $b_{\mu}=0,1$ and
$\gm^{x+b}$ is a short hand notation for the product
$\gm_1^{x_1+b_1} \gm_2^{x_2+b_2} \gm_3^{x_3+b_3} \gm_4^{x_4+b_4}$.
Since $\Psi$ contains 16 times as many degrees of freedom
as $\chi$, not all components $\Psi^{\al\kp}$ are independent.
However, the components of the
low momentum modes $\tilde{\Psi}(p)$ with $-\pi/2<p_{\mu}\leq \pi/2$
are independent \cc{Sm91}.
Spin-flavor transformations on $\chi$ correspond to
discrete transformations on $\Psi$ from the left or right.
A shift transformation $\chi_x \ra \zt_{\rh x} \chi_{x+\rhh}$,
to be interpreted as a discrete flavor transformation \cc{DoSm83,GoSm84},
translates into $\Psi_x \ra \Psi_{x+\rhh}\gm_{\rh}$ and a `spin'
transformation $\chi_x \ra \et_{\rh x} \chi_{x+\rhh}$
translates into $\Psi_x \ra \gm_{\rh} \Psi_{x+\rhh}$.
The sign factors $\et_{\mu x}$ and $\zt_{\mu x}$ are defined as
$\eta_{\mu x}=(-1)^{x_1+ \cdots + x_{\mu-1}}$
and $\zeta_{\mu x}=(-1)^{x_{\mu+1}+\cdots +x_4}$.

The kinetic part of the free staggered fermion action
\be
  S_{K} =  -\half \sum_{x\mu} \et_{\mu x}(\chb_x\chi_{x+\mh}
                      - \chb_{x+\mh}\chi_x)
\ee
can now be written as
\be
  S_{K} =  -\sum_{x\mu} \half \Tr
          ( \Psb_x \gm_{\mu}\Psi_{x+\mh} -
                    \Psb_{x+\mh} \gm_{\mu}\Psi_{x} ),\lb{SFREE}
\ee
which reduces in the classical continuum limit to the gradient term
in eq.~(\ref{SFC}), however with two doublets.

In the reduced staggered formalism the field $\chi$ is restricted
to the odd sites $\chi_x \ra \half (1-\ep_x) \chi_x$
and $\chb_x$ to the even sites $\chb_x \ra \half(1+\ep_x)\chb_x$,
with $\ep_x=(-1)^{x_1+x_2+x_3+x_4}$.
Inserting these restricted fields into (\ref{SDT}) and dropping the
bar on $\chb$ gives
\be
\Psi_x=\eighth \sum_{b}  \gm^{x+b}
\half (1-\ep_{x+b}) \chi_{x+b} \;\;,\;\;\;
\Psb_x=\eighth \sum_{b} (\gm^{x+b})^{\dg}
\half (1+\ep_{x+b})  \chi_{x+b}\lb{SDTR}.
\ee
The action (\ref{SFREE}) with $\Psi$ and $\Psb$
defined as in (\ref{SDTR}), reproduces the kinetic term of the action for
reduced (`real' or `Majorana-like') staggered fermions,
\be
  S_{K} = -\half  \sum_{x \mu} \et_{\mu x} \chi_x \chi_{x+\hmu}\;.
\ee
The restriction of $\chi$ and $\chb$ to odd and even sites
corresponds to the projections
$\Psi \ra \half (\Psi - \gm_5 \Psi \gm_5)$
and $\Psb \ra \half (\Psb + \gm_5\Psb\gm_5)$. More explicitly this
implies that the matrix fields have the structure
\be
      \Psb =    \left( \begin{array}{cc}
                     \psb_L &   0    \\
                        0   & \psb_R \end{array} \right),\;
      \Psi =    \left( \begin{array}{cc}
                        0    & \ps_R \\
                       \ps_L &    0    \end{array} \right).  \;
\ee
The relation of the $2\times 2$ matrix fields $\ps_{L,R}$ and $\psb_{L,R}$
to the fields $\ps$ in the target action (\eq{SFC}) becomes clear when
writing the Yukawa interaction in (\eq{SFC}) in the form
$y (\psb_{L,i \alpha} \ps_{R,
\alpha j} \phi_{ij} + \psb_{R,i \alpha} \ps_{L, \alpha j}
\phi^{\dagger}_{ij})$,
where $\alpha=1,2$ and $i,j=1,2$ are the Weyl spinor and flavor indices,
respectively.

We can also rewrite the Yukawa term
in terms of the matrix fields $\Psi$ and $\Psb$ if we introduce
the $4\times 4$ matrix
\be
      \Phi =    \left( \begin{array}{cc}
                      0      & \ph      \\
                   \ph^{\dg} &  0      \end{array} \right) =
            - \Sm \vr_{\mu}\gm_{\mu}.
                 \lb{DEFPHI}
\ee
The total fermionic action
\be
   S_F     =   -\sum_x [  \Sm \half \Tr
          ( \Psb_x \gm_{\mu}\Psi_{x+\mh} -
                     \Psb_{x+\mh} \gm_{\mu}\Psi_{x} ) +
    y \Tr ( \Psb_x \Psi_x \Phi_x^T) ]
                 \lb{SF}
\ee
reduces in the classical continuum limit to eq.~(\eq{SFC}).
Using (\eq{SDTR}) we finally obtain the action in terms
of the independent $\chi$ fields
\be
    S_F = -\half  \sum_{x \mu} \chi_x \chi_{x+\hmu}
          ( \eta_{\mu x} + y \ep_x \zeta_{\mu x} \vphibar_{\mu \;x} )
        = -\half \sum_{x,y} \chi_x M_{xy} \chi_y , \lb{SCHI}
\ee
where $\vphibar_{\mu \; x} = \frac{1}{16} \sum_b \varphi_{\mu \; x-b}$
is the average of the scalar field over a lattice hypercube.
%
%
The fermion matrix $M$ in eq.~(\eq{SCHI}) is antisymmetric and real.
%

For completeness we also show the action of the Higgs fields,
\be
  S_H = \kp
       \sum_{x \mu} \half \Tr(\ph^{\dg}_x \ph_{x+ \mh}
       + \ph^{\dg}_{x+ \mh} \ph_x)  -
      \sum_x  \half \Tr \left[ \ph_x^{\dg} \ph_x +
      \lm ( \ph_x^{\dg}\ph_x  - \un )^2  \right] \;. \lb{SH}
\ee
The total action $S=S_H + S_F$ depends on three coupling constants,
the Yukawa coupling $y$, the scalar field hopping parameter $\kp$
and the quartic self-coupling $\lm$.

We emphasize that this action is invariant under the staggered
fermion symmetry group:
One can check \cc{BoFr92a} the invariance under shifts,
$\chi_x \ra  \zt_{\rho x} \chi_{x+\rhh} ,\;\;
\varphi_{\mu \; x} \ra \varphi_{\mu \; x+\rhh} (1-2\delta_{\mu\rho})$,
$90^o$ rotations, axis reversal and global U(1) transformations of the form
$\chi_x \ra \exp( i \alpha \ep_x ) \chi_x$ (cf. refs.~\cc{DoSm83,GoSm84}).

The action is not invariant, however, under the full O(4) flavor group,
and one expects to need counterterms to recover this invariance
in the scaling region. In the scaling region operators
with dimension larger than four become irrelevant. There are
two operators with dimension four which respect the discrete
symmetries but break O(4):
\be
O^{(1)}=  \sum_{x \mu} \varphi_{\mu \; x}^4 , \;\;\;\;
O^{(2)}=  \sum_{x \mu} (\varphi_{\mu \; x+\hmu} -\varphi_{\mu \; x})^2
. \lb{C2}
\ee
In general one has to add these operators as counterterms to the
action, $S \ra S + c_1 O^{(1)} + c_2 O^{(2)} $
and tune $c_{1,2}$ as a function of $\kp$ and $y$
in order to recover the O(4) invariance in the scaling region.

Here and also in the numerical work
we will restrict ourselves to the case of radially frozen Higgs fields
corresponding to $\phi \in$SU(2) or equivalently $\lambda = +\infty$.
%
%
%
%
%
%
%
\subsection*{3. Phase diagram and fermion mass}
To investigate the phase diagram
we carry out a simple mean field calculation in
the saddle point formulation
\cc{DrZu83} with the replacement
$\varphi_{\mu \; x} \ra f_{\mu} + \ep_x \fst_{\mu}$.
The constant fields $f_{\mu}$, $\fst_{\mu}$, account for a possible
ferromagnetic and antiferromagnetic ordering
of the scalar field. In this approximation
we find $\vphibar_{\mu \; x} \ra f_{\mu}$ which implies that
the fermion fields in eq.~(\eq{SCHI}) do not couple to the
staggered mode $\fst_{\mu}$.

At $y=0$ the model reduces to the O(4) model which has three
phases: a broken  (or ferromagnetic (FM)) phase for $\kp > \kp_c$
$(\kp_c = 0.30411(1))$, a symmetric (or paramagnetic (PM)) phase
for $-\kp_c < \kp < \kp_c$ and an antiferromagnetic (AM) phase
for $\kp < -\kp_c$. Appropriate order parameters for a distinction
of the various phases are the magnetization
$v_{\mu}=\lag \frac{1}{V} \sum_x \vr_{\mu x} \rag$
and the staggered magnetization
$\vst_{\mu}=\lag \frac{1}{V}\sum_x \ep_x \vr_{\mu x} \rag$ with magnitudes
$v=(\sum_{\mu} v_{\mu}^2)^{1/2}$ and
$\vst=(\sum_{\mu} {\vst_{\mu}}^2)^{1/2}$. In the mean field
approximation $v_{\mu} = f_{\mu}$, $\vst_{\mu}=\fst_{\mu}$.

Since the fermions do not couple to $\fst_{\mu}$
the phase transition between the PM and AM phases comes out
independent of $y$, $\kp_c^{AM-PM}(y)=-1/4$.
For the position of the FM-PM phase transition we find
\be
\kp_c^{FM-PM}(y)  =    \frac{1}{4} - \frac{N_D}{128} y^2
     \sum_p
              \frac{ \sum_{\mu} \cos^2 p_{\mu}  }
                   { \sum_{\mu} \sin^2 p_{\mu} }
     \approx \frac{1}{4}-0.012 N_D y^2  \;,       \lb{PT}
\ee
where the $\sum_p$ is a normalized sum over lattice momenta
$p_{\mu} \in (-\frac{\pi}{2},\frac{\pi}{2}]$.
We have inserted in (\eq{PT}) the number of SU(2)
doublets as the variable $N_D$. In the numerical simulation with
the HMCA, $N_D$ has to be
chosen as a multiple of two in order to guarantee a positive Boltzmann
weight, $[\pm (\Det M )^{1/2}]^{N_D} e^{S_H}>0$. The
mean field results for $\kp_c^{FM-PM}(y)$ and
$\kp_c^{PM-AM}(y)$ ($N_D=2$) are represented by the two dashed curves in
fig.~1. The lines intersect at the point $(\kp,y)\approx (-0.25, 4.67)$.
Thus there is a ferrimagnetic (FI) phase
where both $v_{\mu}$ and $\vst_{\mu}$ are nonzero.

%
%
%
\begin{figure}[tb]
\vspace{8cm}
\caption{ \noindent {\em The phase diagram at $\lambda=\infty$.
The squares represent the transition points determined on an $8^4$ lattice,
the dashed lines are the results of the mean field calculation.}}
\label{F1}
\end{figure}
With the mean field ansatz for $\varphi_{\mu \; x}$
the model describes free fermions and it is
straightforward to compute the fermion propagator.
After Fourier transforming the $\chi$ field in the action (\eq{SCHI})
and following the steps outlined in refs.~\cc{DoSm83,GoSm84} we find
(using $v_{\mu}=f_{\mu}$)
\bea
S_{AB}(p) &=& \lag
     \sum_{x,y} e^{i(p+\pi_A)x}M^{-1}_{xy} e^{-i(p+\pi_B)y} \rag
  / V                                 \lb{PR}       \\
          &\ra& \frac{ -i \sum_{\mu} \Gm_{\mu AB} \sin p_{\mu}
                   -  \sum_{\mu} y v_{\mu} (\Xi_{\mu} \Gm_5 \Xi_5)_{AB}
                      \cos p_{\mu}          }
                   {  \sum_{\mu} \sin^2 p_{\mu} + \sum_{\mu} y^2 v_{\mu}^2
                      \cos^2 p_{\mu}        }  \;, \lb{PROP}
\eea
with $p_{\mu} \in (-\frac{\pi}{2},\frac{\pi}{2}]$
and $\pi_A$ denoting the 16 momentum
four-vectors with components equal to $0$ or $\pi$. The 16 dimensional
gamma and flavor matrices $\Gm_{\mu}$ and $\Xi_{\mu}$
are defined in refs.~\cc{DoSm83,GoSm84}.
For $p \ra 0$ we can read off the fermion mass from eq.~(\ref{PROP}):
$m_F=y v$.
This reproduces the usual tree level relation between $m_F$ and $v$. \\
\subsection*{4. Symmetry breaking terms}
Next we want to estimate the magnitude of the
O(4) symmetry breaking terms which are
induced by the fermions, using renormalized perturbation theory.
To that end we expand $(N_D/2) \mbox{Tr} \ln M$
in powers of the scalar field. From the two-point contribution we find
a contact term $(\dl_R/2) \Sm p_{\mu}^2\vr_{\mu,R}^2$, corresponding to
the counterterm $O^{(2)}$. For the coefficient we obtain
\be
 \dl_R = \frac{N_D}{32} y^2_R \left\{ 1-
     \sum_p
              \frac{ \sum_{\mu} \cos^2 p_{\mu}  }
                   { \sum_{\mu} \sin^2 p_{\mu} + y_R^2 \sum_{\mu} v_{\mu,R}^2
                     \cos^2 p_{\mu}      }  \right\}
     \equiv f_{\dl} N_D y_R^2 \;, \lb{DL}
\ee
where the subscript $R$ indicates a renormalized quantity.
We remark that keeping the term
$y^2_R \sum_{\mu} v^2_{\mu,R} \cos^2 p_{\mu}$
in the fermion loop, gives $O(a^2)$ corrections to $f_{\dl}$
which are negligible deep enough in the scaling region
but which we want to include if $m_F=y_R v_R $ is of order one.

The second term $\vep_R \sum_{\mu}  \vr_{\mu,R}^4$ appears
in the four-point contribution to the effective action. We find,
\be
  \vep_R =  \frac{N_D}{32}y^4_R \sum_p \frac{\sum_{\mu} \cos^4 p_{\mu}
       - \frac{1}{3} \sum_{\mu\not=\nu} \cos^2 p_{\mu}\cos^2 p_{\nu}  }
       { (\sum_{\mu} \sin^2 p_{\mu} + y_R^2 \sum_{\mu} v_{\mu,R}^2
       \cos^2 p_{\mu})^2 }
       \equiv  f_{\vep} N_D y_R^4 \;. \lb{EP}
\ee
%

%
This leads us to
consider the following tree-level effective action,
%
%
%
%
\be
 S_{eff} = -\int d^4x [ \half \sum_{\mu\nu}
 \dm {\vr}_{\nu,R}\dm {\vr}_{\nu,R}(1 + \dl_R \dl_{\mu\nu})
        + \frac{m_R^2}{2} \Sm {\vr}_{\mu,R}^2 + \frac{\lm_R}{4}
          (\Sm {\vr}_{\mu,R}^2)^2
        + \vep_R \Sm {\vr}_{\mu,R}^4 ] \;.        \lb{SR}
\ee
As a consequence of the $\vep_R$-term the shape of
the effective potential differs from the usual Mexican hat form.
The  potential now has 16 discrete minima at
$\vr_R = (\pm\half,\pm\half,\pm\half,\pm\half) v_R$ with
\be
 v_R     = (-m_R^2 /(\lm_R+\vep_R) )^{1/2}. \lb{V}
\ee
In the infinite volume limit the scalar field will be frozen to one of
these minima. Then we can replace
$y_R^2 \sum_{\mu} v_{\mu,R}^2 \cos^2 p_{\mu}$ in eqs.~(\eq{DL}) and (\eq{EP})
by $(m_F^2/4) \sum_{\mu} \cos^2 p_{\mu}$ and
compute the prefactors $f_{\dl}$ and $f_{\vep}$
as a function of $m_F$. We find
$f_{\dl} \approx -0.0150$, $-0.0110$, $-0.0070$ and
$f_{\vep}\approx 0.0054$,
$0.0050$, $0.0043$ for $m_F=0$, $0.3$, $0.5$.

In order to estimate the effect of the symmetry breaking terms on masses
of the longitudinal ($\sigma$) and transversal ($\pi$) modes
we compute now the scalar field propagator for the effective action
in eq.~(\eq{SR}). For this purpose we decompose $\vr$
into its longitudinal and transversal components,
$\vr_{R,\mu}=(v_R+\sigma_R) e^4_{\mu}+ \pi^j_R e^j_{\mu}$,
where $e^4$ is chosen to point in the direction
of one of the 16 minima and $e^j$ are three orthogonal vectors.
We shall make the convenient choice $e^1=\half (1,-1,-1,1)$,
$e^2=\half (-1,1,-1,1)$, $e^3=\half (-1,-1,1,1)$, $e^4=\half (1,1,1,1)$.
In this basis the inverse scalar propagator has the form
\be
G^{-1}(p)^{\alpha \beta} = (1+\frac{\dl_R}{4}) p^2
\dl^{\alpha \beta} + 2 (\lm_R+\vep_R)v_R^2
\dl^{\alpha 4} \dl^{\beta 4} + 2 \vep_R
v_R^2 \dl^{\alpha j} \dl^{\beta j}
+ \dl_R [ \sum_{\mu} p_{\mu}^2 e^{\alpha}_{\mu} e^{\beta}_{\mu}
- \frac{p^2}{4}
\dl^{\alpha \beta} ]    \;.         \lb{INVP}
\ee
Since $\dl_R$ is small compared to $1$ we can
treat the off-diagonal part (i.~e.
the last term in (\eq{INVP})) as a perturbation. Then we find for the
propagators $G_{\sigma}(p)=G^{44}(p)$ and
$G_{\pi}=\frac{1}{3} \sum_j G^{j j}(p)$ the forms
\be
G_{\sigma, \pi}(p)= \frac{ (1+\dl_R/4)^{-1} }
               {p^2 + m_{\sigma,\pi}^2} (1+O(\dl_R^2)) \;, \lb{PROPS}
\ee
where the $\dl^2_R$
correction is bounded by $\frac{3}{4} \dl^2_R$ for all
momenta. The particle masses are given by
\be
m_{\sigma}^2=\frac{2(\lm_R+\vep_R) v_R^2}
                  {1+\dl_R/4}\;,\;\;
m_{\pi}^2=\frac{2 \vep_R v_R^2}
                  {1+\dl_R/4} \;.  \lb{M}
\ee
This shows that the three transversal modes in the scalar spectrum
acquire a mass as a consequence of the
symmetry breaking term in the potential.

We can get an impression of the size
of the symmetry breaking corrections
for $y_R <3$, $N_D=2$ and $m_F=0$. Then $|\dl_R/4|<0.068$,
$\frac{3}{4} \dl^2_R <0.055$ and
$\vep_R < 0.88$ where the latter one has to be compared
with typical values of $\lm_R=5-10$.
The values which we find for these quantities in the physically relevant
region are actually smaller than these bounds.
It should be kept in mind that we have neglected
in the relations (\eq{M}) the
usual loop corrections not involving the symmetry breaking terms.

Note that to leading order in $\dl_R$
the propagators
$G_{\sigma}(p)$ and $G_{\pi}(p)$ in (\eq{PROPS})
are covariant functions of $p_{\mu}$ and $e^4_{\mu}$, such that
we can rotate our frame of reference and
choose for example $e^4=\hat{4}$.
In the following we shall use the approximation
(\eq{PROPS}) for our analysis of the numerical propagator results.
\subsection*{5. Results of the numerical simulations}
In this section we present our results for the phase diagram and give some
preliminary results for the renormalized couplings $y_R$ and $\lm_R$.
The results were obtained at $\lm=\infty$ and
without the two counterterms. The numerical simulations were performed
on $L^3 T$ lattices with periodic boundary conditions in all directions
except for the fermion fields which had antiperiodic boundary conditions
in the time direction.
For the calculation of the fermion and scalar propagators we carried out
the simulations on $L^3 24$ lattices with $L=6,8,10,12$ and $16$ and at
$\kp=0$.
%

The phase diagram is shown in fig.~1. The squares represent the
transition points which were obtained by
scanning systematically the ($\kp,y$) coupling parameter space
in vertical and horizontal directions. For each point in the raster
we have typically accumulated a statistics of
500 scalar field configurations on an $8^4$ lattice.
We used the rotation technique
for the calculation of the order parameters $v$ and $v_{st}$
\cc{BoDe91a,BoDe91b}.
%
%
The numerical results are qualitatively
well described by the mean field results.
The phase diagram in fig.~1
is very similar to the phase diagrams of other models
with a hypercubic Yukawa coupling \cc{HyMo}. 

For the determination of the fermion mass $m_F$ we have
measured the fermion propagator $S_{AB}(p)$ given in (\eq{PR})
for momenta $p=(\vec{0},p_4)$ and computed from it the projections
$S_0 (p_4)=i \mbox{Tr}[ \Gm_4 S(p_4)]/16$,
$S_{\mu} (p_4)=\mbox{Tr}[ \Xi_{\mu} \Xi_5 \Gm_5 S(p_4)]$/16.
The components $S_{\mu}$ ($\mu\not= 0$)  are non-invariant
under cubic rotations of the scalar field
and would vanish when averaging over many
configurations. To avoid finding zero for all of them
we have rotated the scalar fields such that $v_{\mu} = v \delta_{\mu,4}$.
Then, the numerical values for $S_{i}$, $i=1,\ldots,3$ are found to be zero
within errors and we have fitted the non-vanishing
components $S_0$ and $S_4$ to a free fermion propagator
form (cf. (\ref{PROP}))
%
%
\be
S_0(p_4) \ra
     \frac{Z_{F,0} \sin p_4}{ (1-m^2_{F,0}) \sin^2 p_4 +m_{F,0}^2} ,\;\;
S_4(p_4) \ra  \frac{Z_{F,4} m_{F,4} \cos p_4}
{ (1-m^2_{F,4}) \sin^2 p_4 +m_{F,4}^2} ,  \lb{FIT}
\ee
where we allow  for two different
masses and wave-function renormalization constants in $S_0$ and
$S_4$. In fig.~2 we have plotted $S_0^{-1}(p_4) \sin p_4$
and $S_4^{-1}(p_4) \cos p_4$ for various values of $y$
as a function of $\sin^2 p_4$.
The data points fall nicely on straight lines
which were obtained by fitting $S_0$ and $S_4$ to
the forms (\eq{FIT}). Within the
statistical errors, we find that $m_{F,0} \approx m_{F,4}$. We therefore have
taken the average, which will be denoted by $m_F$ in the following.
The results for $m_F$ are summarized in table 1
for several $y$ values and several lattice sizes.
We find $Z_{F,0} \approx 0.8$ and
$Z_{F,4}$ approximately 5\% smaller than $Z_{F,0}$.
When rotating $v_{\mu}$ in a different direction we find that
$m_{F,0}$, $Z_{F,0}$ and $m_{F,4}$, $Z_{F,4}$ may differ by about 7\% for
$y<4.2$.
Such systematic effects are presumably due to scaling violations.

%
%
%
\begin{figure}[tb]
\vspace{10cm}
\caption{ \noindent {\em $S_0 (p_4) \sin p_4$ and $S_4 (p_4) \cos p_4$ as
a function of $\sin^2 p_4$ for several values of $y$ ($\kp=0$,
$V=12^324$). The error bars are
in all cases much smaller than the symbols. The
straight lines were obtained by fitting
$S_0(p_4)$ and $S_4(p_4)$ to the forms
({\protect \eq{FIT}}).}}
\label{F2}
\end{figure}
In order to monitor finite size effects and extrapolate to infinite volume,
we have computed $m_F$ and $v$ on lattices of different spatial extent $L$.
In a previous work with naive fermions both observables were found to
obey the relations $v_{L}=v_{\infty}+a_1 /L^2$ and
$m_{F,L}=m_{F,\infty}+a_2 /L^2$ with $a_1,a_2 >0$ \cc{BoDe91b}.
This finite size dependence of $v$ and $m_F$ is due to the massless
Goldstone bosons. In our model
the three transversal modes acquire a mass $m_{\pi}>0$ as a
consequence of the O(4) symmetry breaking and we expect deviations from the
$1/L^2$ behavior if $L$ becomes larger than $O(1/m_{\pi})$.
In fig.~3 we have displayed the $1/L^2$ dependence of $v$ and $m_F$.
The plot shows that the $1/L^2$
dependence is well fulfilled for $y=3.8$ though small deviations
are visible for $m_F$ on the
largest lattice with $L=16$. The deviations become more pronounced
at $y=4.0$, which is consistent with the observed increase of $m_{\pi}$ (see
fig.~4). The effect of the symmetry breaking on $v$ and $m_F$, as measured
by the deviations from the straight line behavior, appears to be smaller
than 15\%.
The effect on the ratio $m_F/v$ is even smaller.

%
%
%
%
%
\begin{figure}[tb]
\vspace{10cm}
\caption{ \noindent {\em The results for
$v$ and $m_F$ as a function of $1/L^2$ for
$y=3.8$ and $y=4.0$. The straight lines are drawn to make deviations
from a linear $1/L^2$ dependence visible.}}
\label{F3}
\end{figure}
The masses $m_{\sigma}$ and $m_{\pi}$
of the longitudinal and transversal modes were obtained
by fitting the scalar momentum space propagators
$G_{\sigma} (p)=\lag \sum_{x,y} \sigma_x \sigma_y e^{ i p (x-y)}
\rag /V$
and $G_{\pi} (p)= \third \sum_j \lag \sum_{x,y}
\pi_x^j \pi_y^j e^{ i p (x-y)}\rag /V$
for sufficiently small momenta to free
boson propagators, cf. (\ref{PROPS}),
\be
G^{-1}_{\sigma,\pi}(p) \ra  Z_{\sigma,\pi}^{-1}
(m_{\sigma,\pi}^2 + \phat ) \;,\;\; p \neq 0 \;. \lb{PROPSS}
\ee
%
%
%
%
%
The quantity $\phat=2\sum_{\mu}
(1-\cos p_{\mu})$ is a lattice equivalent of the momentum squared in the
continuum. Here we have chosen the magnetization
to point in the four-direction as for the computation of the fermion
mass (see also the remarks at the end of sect.~4).
When plotting the inverse propagator $G_{\sigma,\pi}^{-1}(p)$
as a function of $\phat$ we expect to find a linear behavior if $\phat \ll
m_F^2$, as is found in the O(4) model. We measured
the $\sigma$ and $\pi$ propagators on a $12^324$
lattice where $G_{\sigma,\pi}^{-1}(p)$ for
the three smallest momenta has an
approximately linear behavior. We note that because of curvature in
$G_{\sigma,\pi}^{-1}(p)$, the results of a linear fit tend to overestimate
$m_{\sigma,\pi}$ and $Z_{\sigma,\pi}$.
In a forthcoming publication \cc{BoFr92a} we will extend the analysis of the
scalar propagators by taking into account the one-fermion-loop
contribution to the self-energy which allows for a good description of the
curvature in $G_{\sigma , \pi}^{-1}(p)$.

In fig.~4 we have plotted the fit results for $m_{\pi}$ (squares)
as a function of $y$. The mass has a minimum close
to $y=3.8$ which is due to two different effects. The increase
towards small $y$ is a consequence of the finite lattice size as
in the pure O(4) theory. The mass
has to approach the scalar mass $m_S$ in the PM phase
continuously when crossing the FM-PM phase boundary region.
On a finite lattice $m_S$ does not vanish at the FM-PM phase transition and
consequently $m_{\pi}$ has to grow when approaching the FM-PM phase transition.
The increase at large $y$ is due to the O(4)
symmetry breaking. The numerical
data for $m_{\pi}$ may be compared with the one-loop results given in
eqs.~(\eq{EP}) and (\eq{M}). The values obtained for $m_{\pi}$
are represented in fig.~4 by the diamonds. For $y \gsim 4.0$
the agreement with the numerical results is reasonable.

%
%
%
%
%
\begin{figure}[tb]
\vspace{8cm}
\caption{ \noindent {\em Numerical results for
$m_{\pi}$ (squares) as a function of $y$
($\kp=0$, $V=12^3 24$). The diamonds
represent the values for $m_{\pi}$ which were obtained by inserting
the numerical results for $m_F$, $y_R$ and $v_R$
into eqs.~({\protect \eq{EP}}) and
({\protect \eq{M}}).}}
\label{F4}
\end{figure}
We will present now some preliminary results for the
renormalized couplings
$y_R$ and $\lambda_R$, using the tree level motivated definitions
$y_R=m_F/v_R$ and $\lm_R=\half (1+\dl_R/4) (m_{\sigma}/v_R)^2-\vep_R$.
The definition of $\lm_R$
is based on eq.~(\eq{M}) and we shall use 
the one-loop values given in
eqs.~(\eq{DL}) and (\eq{EP}) for 
$\dl_R$ and $\vep_R$.
For the normalization of the scalar
fields we use the wave-function renormalization constant $Z_{\pi}$ (not
$Z_{\sigma}$ because the $\sigma$ particle is unstable).
The renormalized field expectation
value is then defined by $v_R=v/\sqrt{ Z_{\pi} }$. We find that
$Z_{\pi}$ is smaller than in the pure O(4) model by roughly a factor four.
In table 1 we give the results for $y_R$, $\sqrt{2 \lm_R}$ and the
ratio $m_{\sigma}/v_R$.
The table shows that the numerical values for
$\sqrt{2 \lm_R}$ and the ratio $m_{\sigma}/v_R$,
which differ by the $\vep_R$ and $\dl_R$ corrections,
agree within 5\%.
This indicates that the effect of the symmetry breaking is
relatively small.

The bare Yukawa couplings are relatively large and
it is therefore interesting to compare the
values for $y_R$ listed in table 1
with the tree level unitarity bound which for this model is given
by $y_R^{u.b.}=2 \sqrt{\pi/N_D} \approx 2.51$ \cc{BoDe91a}.
At the edge of the scaling region with $m_{\sigma} \approx 0.75$
the numerical results for $y_R$ are quite close to this
value. This indicates that the renormalized
couplings are relatively weak, but
they appear to be stronger than in the model
with naive fermions where the
numerical results for $y_R$ are significantly
smaller than $y_R^{u.b.}$ \cc{BoDe91b}.

The values for the quantity $\sqrt{2\lm_R}$ with $L=12$
may be compared with the infinite
volume results obtained previously in the O(4) model \cc{O4}.
There the ratio $\sqrt{2\lm_{R,\infty}}=m_{\sigma,\infty}/v_{R,\infty}$
ranges from 2.5 to 3.1 for $m_{\sigma,\infty}=0.4-1.0$. The finite volume
results in table 1 with $m_{\sigma,L} < 1$
lie clearly above this range which indicates that also the upper bound
of the Higgs mass grows when the Yukawa coupling is turned on.
\begin{table}
\begin{center}
\begin{tabular}{|l|l|l|l|l|l|l|l|l|l|} \hline
$y$ & $L$ & $m_F$ & $v$ & $Z_{\pi}$ & $m_{\pi}$
& $m_{\sigma}$ & $y_R$ & $m_{\sigma}/v_R$ & $\sqrt{2\lambda_R}$ \\
                       \hline\hline
$3.6$ & $12$ & $0.183(10)$ & $0.0443(9)$ & $0.39(1)$ & $0.26(1)$ &
$0.52(22)$ & $2.57(15)$ & $7.3(3.1)$ & $7.1(3.2)$ \\ \hline
$3.8$ & $6$ &  $0.472(15)$ & $0.1095(11)$ &  &  &  &
&   &    \\
$3.8$ & $8$ &  $0.387(9)$ & $0.0931(8)$ &  &  & &
&   &    \\
$3.8$ & $10$ &  $0.341(6)$ & $0.0842(9)$ &  &  & &
&   &    \\
$3.8$ & $12$ & $0.327(4)$ & $0.0796(10)$ & $0.41(2)$ & $0.15(1)$ &
$0.54(11)$ & $2.62(8)$ & $4.3(9)$ & $4.1(9)$ \\
$3.8$ & $16$ &  $0.318(2)$ & $0.0766(10)$ &  &  &  &
&   &    \\  \hline
$4.0$ & $6$ &  $0.575(11)$ & $0.1287(10)$ &  &  & &
&   &    \\
$4.0$ & $8$ &  $0.503(6)$ & $0.1190(5)$ &  &  &  &
&   &    \\
$4.0$ & $10$ &  $0.488(4)$ & $0.1164(7)$ &  &  &  &
&   &    \\
$4.0$ & $12$ & $0.479(3)$ & $0.1154(7)$ & $0.47(2)$ & $0.18(3)$ &
$0.74(7)$ & $2.83(5)$ & $4.4(5)$ & $4.2(5)$ \\
$4.0$ & $16$ &  $0.477(3)$ & $0.1153(6)$ &  &  & &
&   &    \\ \hline
$4.2$ & $6$ &  $0.597(9)$ & $0.1465(13)$ &  &  & &
&   &    \\
$4.2$ & $12$ & $0.603(4)$ & $0.1466(7)$ & $0.51(3)$ & $0.21(2)$ &
$1.16(5)$ & $3.04(9)$ & $5.6(3)$ & $5.4(3)$ \\ \hline
$4.8$ & $12$ & $0.985(3)$ & $0.2156(3)$ & $0.64(2)$ & $0.35(2)$ &
$1.60(8)$ & $3.64(4)$ & $5.9(1)$ & $5.8(1)$ \\ \hline
\end{tabular}
\caption{{ \em Some results for $m_F$, $v$, $Z_{\pi}$, $m_{\pi}$,
$m_{\sigma}$, $y_R$, $m_{\sigma}/v_R$ and $( 2 \lambda_R )^{1/2}$.
The value of $\kp$ is fixed
to $0$. The lattice size is $L^3T$ with $T=24$. }}
\end{center}
\end{table}
\subsection*{6. Conclusion}
We have introduced a fermion-Higgs model with reduced staggered fermions
in which the {\em staggered} flavors are coupled to the Higgs field.
In a numerical simulation with the hybrid Monte Carlo algorithm
the model contains two isospin doublets in the scaling region.
To recover the full O(4) symmetry
we have to add two counterterms to the scalar part of the
action. The addition of counterterms can perhaps be avoided
using an $F_4$ lattice.

In this paper we have studied the model without these
counterterms.
We find that a one-loop computation of the effect of the symmetry
breaking can account reasonably well for the measured values of $m_\pi$
and predicts a 5\% correction in the relation between $m_{\sg}/v_R$ and
$\sqrt{2 \lm_R}$. Also the finite volume dependence of $v$ and $m_F$ indicates
that the effect of the symmetry breaking is relatively small.

As a preliminary result we find
that at $\kp=0$ and at relatively large bare Yukawa coupling $y$ the
renormalized Yukawa coupling cannot be significantly larger than the
tree level unitarity bound in the
region with $m_{\sigma} < 0.7$, $m_F < 0.5$.
The quantity $\sqrt{2 \lm_R}$ comes out to be
larger than the infinite volume results in the pure O(4) model.

The model has proven to be very efficient in a numerical
simulation and it is possible to perform calculations
on relatively large lattices.
The time for an update of a scalar field configuration
is smaller by roughly a factor 10 than in
a model with naive fermions \cc{BoDe91a,BoDe91b}.

We have investigated here a simple example in a
class of models in which the staggered flavors are coupled
to scalar or gauge degrees of freedom \cc{Sm91,Sm88}.
We will investigate in future more complicated models which might provide
a non-perturbative formulation of a chiral gauge theory on the lattice. \\

\noindent {\bf Acknowledgements}.
The numerical calculations were performed on the CRAY Y-MP4/464
at SARA, Amsterdam. This research was supported by the ``Stichting voor
Fun\-da\-men\-teel On\-der\-zoek der Materie (FOM)''
and by the ``Stichting Nationale Computer Faciliteiten'' (NCF).
%
%

%
\end{document}